\documentclass[fleqn,usenatbib]{mnras}

\usepackage{newtxtext,newtxmath}

\usepackage[T1]{fontenc}
\usepackage{ae,aecompl}

\usepackage{graphicx}	
\usepackage{amsmath}	
\usepackage{amssymb}	

\newcommand{\hi}{H\,\textsc{i}}
\newcommand{\hii}{H\,\textsc{ii}}

\title[Cold HI ejected into the MS]{Cold HI ejected into the Magellanic Stream}

\author[Dempsey et al.]{
J.~Dempsey,$^{1,2}$\thanks{E-mail: james.dempsey@anu.edu.au}
N.~M.~McClure-Griffiths$^{1}$,
K.~Jameson$^{1,3}$,
and F.~Buckland-Willis$^{1}$
\\
$^1$ Research School of Astronomy \& Astrophysics, Australian National University, Canberra, ACT 2611, Australia \\
$^2$ CSIRO Information Management and Technology, GPO Box 1700 Canberra, ACT 2601, Australia \\
$^3$ CSIRO Astronomy and Space Science, 26 Dick Perry Avenue, Kensington, WA 6151, Australia
}

\date{Accepted 2020 June 02. Received 2020 June 02; in original form 2019 October 16}

\pubyear{2020}

\begin{document}
\label{firstpage}
\pagerange{\pageref{firstpage}--\pageref{lastpage}}
\maketitle

\begin{abstract}
We report the direct detection of cold \hi\ gas in a cloud ejected from the Small Magellanic Cloud (SMC) towards the Magellanic Stream. 
The cloud is part of a fragmented shell of \hi\ gas on the outskirts of the SMC.
This is the second direct detection of cold \hi\ associated with the Magellanic Stream using absorption.
The cold gas was detected using 21-cm \hi\ absorption-line observations with the Australia Telescope Compact Array (ATCA) towards the extra-galactic source PMN~J0029$-$7228.
We find a spin (excitation) temperature for the gas of $68 \pm 20$ K. 
We suggest that breaking super shells from the Magellanic Clouds may be a source of cold gas to supply the rest of the Magellanic Stream.
\end{abstract}




\section{Introduction}
\label{sec:introduction}

The Small (SMC) and Large Magellanic Clouds (LMC) are the most prominent parts of an interacting galaxy system stretching 200$^\circ$ across the sky, including the Magellanic Bridge between the Magellanic Clouds (MCs), the Leading Arm in front of the MCs and the 125$^\circ$ long Magellanic Stream (MS) trailing behind the MCs \citep{2010ApJ...723.1618N}.
The MS consists of low metallicity, low density gas stripped from the SMC and LMC.
It has little dust \citep{1987MNRAS.224.1059F,2013ApJ...772..110F} and no stars \citep{2016ARA&A..54..363D}.
Two filaments have been traced, predominantly in neutral hydrogen (\hi) observations, through the MS \citep{2003ApJ...586..170P,2008ApJ...679..432N}.
The higher metallicity filament (${\sim}0.50$ solar; \citealt{2013ApJ...772..111R}) traces back to the LMC , while the lower metallicity filament (${\sim}0.1$ solar; \citealt{2013ApJ...772..110F}) extends back to the SMC \citep{2008ApJ...679..432N}.

In this study we focus on the neutral hydrogen in the SMC and the MS.
Most neutral hydrogen exists in two long-lived phases, the Warm Neutral Medium (WNM), with a kinetic temperature ($T_{\rm k}$) of $5,000 \leq T_{\rm k} \leq  8,000$ K, and the Cold Neutral Medium (CNM), with temperatures of $20 \leq T_{\rm k} \leq 200$ K \citep{1977ApJ...218..148M}.
The WNM and CNM are generally found in a pressure equilibrium that can only exist in a narrow range of thermal pressure, which itself is governed by the balance of heating and cooling mechanisms in the gas \citep{1995ApJ...443..152W,2003ApJ...587..278W}.
The dominant heating mechanism is photoelectric heating from dust grains such as Polycyclic Aromatic Hydrocarbons (PAHs) excited by far ultraviolet radiation \citep{1995ApJ...443..152W}. 
In environments with low metallicity or dust ratios, X-ray and cosmic ray heating plays a larger role.
The primary cooling mechanism for gas $T_{\rm k} \lesssim 8000$ K is via the fine-structure C~\textsc{ii} and  O~\textsc{i} lines \citep{1988gera.book...95K,1995ApJ...443..152W}. 
This metal line cooling is less efficient in low metallicity environments and the two-phase equilibrium requires a higher thermal pressure \citep{1995ApJ...443..152W}.
Moderate levels of turbulence can also drive the formation of cold gas \citep{1999A&A...351..309H,2005A&A...433....1A}.

The interstellar medium (ISM) of the SMC is a multi-phase environment.
The main body of the SMC has $\approx~4.0\times10^8$ M$_\odot$ of \hi\ \citep{2005A&A...432...45B}.
There are also large reserves of warm \hii\ at $T_{\rm k} \sim 10^4$ K \citep{2015ASPC..491..343W}.
The SMC is surrounded by a halo of warm \hii\ \citep{2019ApJ...887...16S} and some hot \hii\ at $T_{\rm k} \sim 10^5$ K \citep{2002ApJ...569..233H}.
Whilst the SMC has a metallicity of just $\approx 0.2$ solar \citep{1992ApJ...384..508R}, \cite{2013ApJ...771..132B} found that the dominant heating mechanism is far ultraviolet radiation from the SMC stellar population.
Cooling in the SMC will be inefficient due to the lack of metals.
Outside of the main body of the SMC, the lifetime of cold ($T_{\rm k} \sim 100$ K) \hi\ clouds will be very short unless they are shielded by significant outer \hi\ layers \citep{1977ApJ...211..135C}.

Simulations of dwarf galaxies similar to the SMC show that cool ($T_{\rm k}$ $\lesssim 2000$ K) gas should be present in wind driven outflows \citep{2012MNRAS.421.3522H,2013MNRAS.430.1901H}.
Outflows of cold \hi\ gas \citep{2018NatAs...2..901M} and molecular H$_2$ gas 
\citep{2019ApJ...885L..32D} 
from the SMC have been observed in emission.
However, to directly detect the cold \hi\ ($T_{\rm k}$ $\lesssim 100$ K) we need to turn to 21-cm absorption.
Absorption studies, when combined with emission data, allow us to measure the spin ($T_{\rm S}$) or excitation temperature of the gas. 
In the CNM, $T_{\rm S} = T_{\rm k}$, allowing us to measure the kinetic temperature of the gas, whereas with just emission data we can only provide an upper limit on the kinetic temperature \citep{2001A&A...371..698L}.

In this study we set out to directly detect cold HI gas potentially feeding into the MS using 21-cm absorption. 
Up to 27\% of the total \hi\ mass observed in emission in parts of the MS is cold \citep{2006A&A...455..481K} and cold \hi\ has been directly detected at one location in the MS \citep{2009ApJ...691L.115M}.
However, with significant ionised gas and low metallicity, it is an environment where it is difficult to form cold HI. 
\cite{Buckland-Willis:anu} examined absorption towards background sources on the edges of the SMC with 3.6 km s$^{-1}$ velocity resolution Australian Square Kilometre Array Pathfinder (ASKAP) observations \citep{2018NatAs...2..901M}.
While they identified candidate areas of higher opacity, the velocity resolution was not sufficient to allow quantitative analysis of the absorption properties of the cold \hi, which has a typical line width  $< 3$ km s$^{-1}$.
We chose five of these background sources which probed locations where cold \hi\ gas could potentially feed into the MS, and made high sensitivity absorption observations of them using the Australia Telescope Compact Array (ATCA).

In this paper, we present our \hi\ absorption observations of the five fields, all located on the outskirts of the SMC (see Fig.~\ref{fig:sources}). 
We describe our observations and initial data reduction in section \ref{sec:observations}. 
In section \ref{sec:results}, we describe the analysis of the observations and the results obtained. 
We discuss where cold gas was detected and the implications for future studies in section~\ref{sec:discussion} and 
summarise our findings in section~\ref{sec:conclusions}.

\section{Observations}
\label{sec:observations}

\begin{figure}
	\includegraphics[width=\columnwidth]{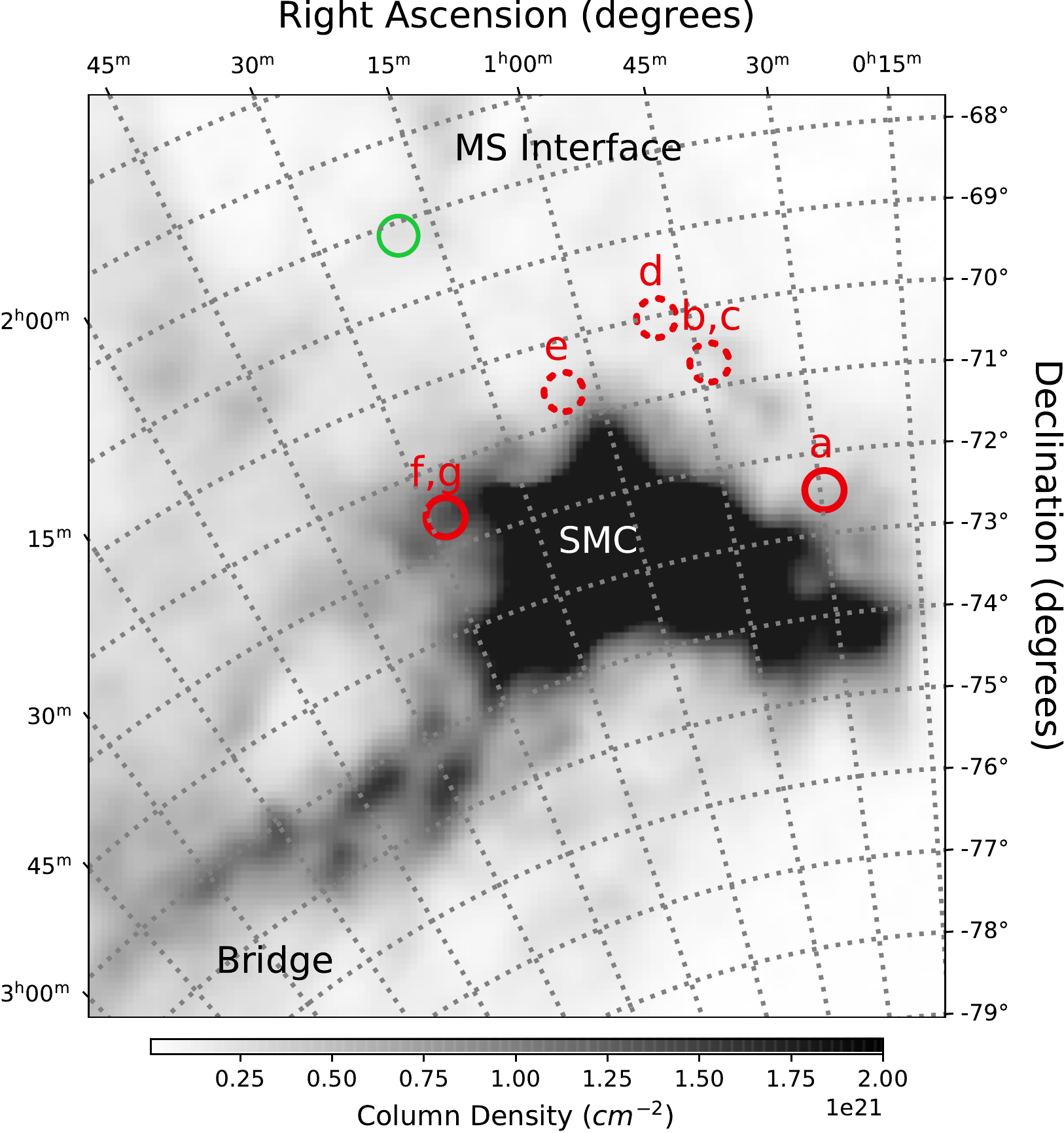}
    \caption{
    Map of our \hi\ targets, shown as red circles. 
    Solid circles indicate locations at which significant absorption was detected, dashed showing location of non-detection or marginal detection.
    The \protect\cite{2009ApJ...691L.115M} detection is shown as a green circle.
    The background shows \hi\ column density from GASS III \citep{2015A&A...578A..78K} with the SMC on the right, the MS interface region on the top and the Magellanic Bridge on the bottom left.
    }
    \label{fig:sources}
\end{figure}

We selected five sources from the \cite{Buckland-Willis:anu} study located on the north and west sides of the SMC (i.e. adjacent to the MS interface region). 
In Fig.~\ref{fig:sources} we show the positions of the selected sources.
\cite{Buckland-Willis:anu} observed flux values for these sources ranging from 31 mJy beam$^{-1}$ (source SUMSS~J012734$-$713639A/B) to 388 mJy beam$^{-1}$ (source ATPMN~J004330.6$-$704148) with a 35 x 27 arcsec$^2$ beam.

We took observations between 24 May 2019 and 2 June 2019 (project code C3297) using the ATCA with the same observational approach as \cite{2019ApJS..244....7J}.
We summarise this approach below.
We observed each source 20 times in 30 minute integrations spread over 11 hours for a total of 600 minutes.
Between each integration, we observed the phase calibrator source PKS 0252$-$712 for two minutes.
We observed the standard ATCA primary calibrator source, PKS 1934$-$638, for 50 minutes at the start of each session for flux and bandpass calibration.
All observations were taken using the 6A (6 km baseline) configuration of the ATCA, giving a beam size of 12.3 x 7.5 arcsec$^2$ at 1420 MHz.
We used the 1M-0.5k configuration of the Compact Array Broad-band Backend (CABB; \cite{2011MNRAS.416..832W}) to observe both a 2GHz continuum band centred on 2100 MHz and a 1 MHz range at high spectral resolution across the \hi\ 21-cm band.
For the \hi\ band, this resulted in a kinematic Local Standard of Rest (LSRK) velocity coverage of -100 $\leq v_{\rm LSRK} \leq$ 500 km s$^{-1}$ with a 0.1 km s$^{-1}$ spectral resolution.
For a detailed description of the observing parameters, please see \cite{2019ApJS..244....7J}.

We flagged RFI and bad channels, then calibrated and imaged the data using standard techniques in Miriad \citep{1995ASPC...77..433S}.
As part of the imaging process, we averaged the \hi\ data from the original 0.1 km s$^{-1}$ to 1 km s$^{-1}$ spectral resolution to reduce the noise level. 
We then used the Miriad \textsc{selfcal} task to perform self-calibration on the fields in order to further reduce noise levels.
This used the model of the sources in the field, as produced by the earlier deconvolution steps, as the input to refine the calibration of the antenna phase gains over time.
For each field, we produced an \hi\ cube and a multi-frequency synthesis (MFS) continuum image. 
The \hi\ cubes have a mean noise level per 1 km s$^{-1}$ channel of 3.25 mJy beam$^{-1}$ and a velocity range of -90 $\leq $ v$_{\rm LSRK}$ $ \leq$ 476 km s$^{-1}$.
The signal-to-noise ratio was $5.5 \leq$ S/N $\leq 27$  for our target sources (see Table~\ref{tab:sources_table}).
Flux values observed in this study for the sources 0127$-$7136 and 0043$-$7041 were smaller than those in \cite{Buckland-Willis:anu} as these sources were resolved in the smaller ATCA beam.

\section{Results}
\label{sec:results}

\begin{table*}
\centering
\caption{List of observed sources}
\label{tab:sources_table}
\begin{tabular}{llrrrrrrrrr}
\hline
 Source & Source Name & RA & Dec & $S_{\rm cont}$ & Peak $\tau$ & v$_{\rm LSRK}$  & Peak $T_{\rm B}$ &  $N_{\rm \hi}$ \\
 &  &  &  &  &  & Peak $\tau$ &   & ($10^{20}$ \\
 &  & (deg) & (deg) & (mJy) &  & (km s$^{-1}$) & ($\mathrm{K}$) & cm$^{-2}$)\\
\hline
a) 002919$-$722811 & PMN~J0029$-$7228 & 7.3288 & -72.470 & 88 & $0.21 \pm 0.05$ & 123 & $12.9\pm 2.6$ & 6.3 \\
b) 004327$-$704136A & ATPMN~J004326.9$-$704135 & 10.8626 & -70.693 & 70 & $<0.07 \pm 0.02$ & 78 & $2.5\pm 1.1$ & 1.4 \\
c) 004327$-$704136B & ATPMN~J004330.6$-$704148 & 10.8781 & -70.697 & 74 & $<0.07 \pm 0.03$ & 100 & $2.3\pm 0.8$ & 1.3 \\
d) 004938$-$700133 & PMN~J0049$-$7001 & 12.4068 & -70.026 & 57 & $<0.15 \pm 0.06$ & 128 & $1.6\pm 0.6$ & 1.1 \\
e) 010458$-$703735 & PMN~J0104$-$7038 & 16.2420 & -70.626 & 27 & $<0.35 \pm 0.17$ & 184 & $4.4\pm 0.8$ & 2.7 \\
f) 012734$-$713638A & SUMSS~J012734$-$713639A & 21.8897 & -71.610 & 18 & $0.45 \pm 0.17$ & 204 & $32.2\pm 1.9$ & 15.8 \\
g) 012734$-$713638B & SUMSS~J012734$-$713639B & 21.8923 & -71.611 & 20 & $0.42 \pm 0.18$ & 196 & $31.8\pm 2.0$ & 15.7 \\
\hline
\end{tabular}
~\\
Sources are from PMN: \cite{1994ApJS...91..111W}, ATPMN: \cite{2012MNRAS.422.1527M} and SUMSS: \cite{2003MNRAS.342.1117M}
\end{table*}

We used the \textsc{Aegean} \citep{2012MNRAS.422.1812H,2018PASA...35...11H} source finding package to analyse the continuum images and characterise the sources.
We configured \textsc{Aegean} with a 5$\sigma$ signal-to-noise ratio (S/N) threshold for source detection and a 4$\sigma$ S/N threshold for adjacent pixels to be part of the same island.
We filtered the detected sources, requiring S/N $\geq 10$ and $S_{\rm cont} \geq 15$ mJy, thus excluding residual sidelobes and sources where we were unlikely to be able to produce a reasonable signal-to-noise spectrum from the \hi\ data. 
In all fields observed, only the central source was included based on this filter.
However in two cases the central targets were identified as double sources at high angular resolution.
The central source in field 0043$-$7041 is a resolved pair with clear separation between the components.
In field 0127$-$7136 the central source is elongated with two peaks and a dimming between the peaks, indicative of an unresolved pair (see Fig.~\ref{fig:mfs-day5}).  
The resulting seven sources are listed in Table~\ref{tab:sources_table}.

To extract the absorption spectra, we used the spatial pixels of the \hi\ cube which were inside the source ellipse, as identified by \textsc{Aegean}. 
We then measured the continuum level by using the mean values of the spatial pixels in the velocity range of $-80 \leq v_{\rm LSRK} \leq -20$ km s$^{-1}$, where no line emission was measured in the GASS III \citep{2009ApJS..181..398M,2015A&A...578A..78K} data.
Lastly, we used the continuum level of each pixel to weight the contribution of the pixel to the overall spectrum using \cite[eq. 2]{1992ApJ...385..501D}:
\begin{equation}
 e^{-\tau(v)}  = \sum_{\rm i} \left[ \left( \frac{c_{\rm i}^2}{\sum_{\rm j} c_{\rm j}^2} \right) s_{\rm i}(v)\right]\ ,
\label{eq:3-weighting}
\end{equation}
where $c_{\rm i}$ is the continuum flux for the $i$th pixel and $s_{\rm i}(v)$ is the flux of the pixel at the velocity step.
An example of a resulting spectrum is shown in Fig.~\ref{fig:spectrum-day3}.

We summarise our sources in Table~\ref{tab:sources_table}.
The RA, Dec and brightness (S$_{\rm cont}$) of the sources are all measured from the continuum image.
The peak optical depth (Peak $\tau$) and LSRK velocity of Peak $\tau$ are measured from the absorption spectrum.
The peak brightness temperature ($T_{\rm B}$) is measured from the ASKAP emission spectrum within the SMC velocity range ($75 \leq$ v$_{\rm LSRK}$ $\leq 275$).
The column density (N$_{\rm \hi}$) is calculated, under the assumption that the \hi\ gas is optically thin ($\tau \ll 1$), using the integral of the emission spectrum over the SMC velocity range \citep[eq. 3]{1990ARA&A..28..215D}:

\begin{equation}
    N_{\rm \hi} = 1.823 \times 10^{18} \int T_{\rm B}(\text{v})~d\text{v}~\text{cm}^{-2}
\end{equation}
We matched the sources to their closest known appropriate object in NED and SIMBAD and recorded this in the source name column.

Two background sources, PMN J0029$-$7228, and SUMSS~J012734$-$713639A/B, had clear absorption features, as shown in Figures~\ref{fig:spectrum-day3} and \ref{fig:spectrum-day5}. 
Each is described in its own subsection below.
The other four sources listed in Table~\ref{tab:sources_table}, including another source pair, did not have absorption with more than one consecutive channel above 2$\sigma$ significance and thus any potential absorption could not be distinguished from noise.  

We used the combined data cube of continuum subtracted ASKAP plus Parkes observations of the SMC \citep{2018NatAs...2..901M} to measure the emission. 
To estimate the emission at the location of each source, we took 18 single pixel samples in a circle 2 arcmin from the source position, as shown in Fig.~\ref{fig:emission-day3}.
The mean of the spectra from each sample was taken as the emission spectrum at the point, and the $1\sigma$ noise level measured from the samples.
We then linearly interpolated the spectra to match the 1 km s$^{-1}$ scale of the absorption spectra.
The noise levels in the emission spectra are dominated by variations between the samples, particularly when sampling near the edge of a small cloud.
Line emission will increase the antenna temperature of the ATCA observations despite the long baselines filtering out this large scale emission.
As a result we have scaled the noise in each channel of the absorption spectrum by the increase in antenna temperature from emission in that channel \citep{2019ApJS..244....7J}.

We decomposed each emission spectrum into a set of Gaussians using the \textsc{GaussPy+} automated Gaussian decomposition tool \citep{2019A&A...628A..78R,2015AJ....149..138L}.
We also used \textsc{GaussPy+} to decompose those absorption spectra with a per-channel signal-to-noise ratio $S/N > 3$.
Whilst we could have done this by hand for the small number of spectra here, we chose to use a fully autonomous fitting routine as a trial for the upcoming Galactic ASKAP (GASKAP; \citealt{2013PASA...30....3D}) survey.
In such a large survey, with over 100,000 \hi\ absorption spectra expected, manual fitting is impractical.
\textsc{GaussPy+} determines the initial number and mean velocity of Gaussian components using the derivatives of the spectrum.
It then iteratively refines the components using the $\chi^2$ fit and adjusting the number of Gaussians.
We used the default configuration of \textsc{GaussPy+}, as detailed in \cite{2019A&A...628A..78R}, with only two changes:
1) For the absorption spectra, we set the signal-to-noise threshold to 3$\sigma$;
2) For the emission spectra we set a maximum FWHM (the line's full width at half maximum) of 250 km s$^{-1}$ to exclude non-physical wide components.

We estimated the spin temperature of the gas by comparing the observed emission and absorption for a source where both were detected. 
For source PMN J0029$-$7228, with detected emission and absorption, we used the approach described by \cite{2017ApJ...837...55M} to estimate the spin temperature of the gas.
We calculated the isothermal spin temperature at each velocity step of the component as $T_{\rm S}(v) = T_{\rm B}(v) / (1-exp(-\tau(v)))$ \citep[Eq 12]{2017ApJ...837...55M}.
The isothermal spin temperature is the spin temperature at a velocity channel if a single \hi\ component was producing both the emission and absorption at that velocity channel.
We then calculated the optical depth-weighted spin temperature for the absorption component using \citep[Eq 13]{2017ApJ...837...55M}
\begin{equation}
    T_{\rm S} = \frac{\int \tau(v) T_{\rm S}(v) \textrm{d}v}{\int \tau(v)\textrm{d}v}
    \label{eq:spin-temp}
\end{equation}
For the other sources, where the absorption could not be fitted, we estimated the spin temperature by comparing the peak tau with the brightness temperature at the same velocity, 
\begin{equation}
    T_{\rm S} \approx \frac{T_{\rm B}(v)}{1-e^{-\tau(v)}}
    \label{eq:spin-temp-est}
\end{equation}

We can calculate an upper limit on the kinetic temperature of a cloud, using 
T$_{\rm k,max} = 21.86 \Delta V^2$ \citep{2003ApJS..145..329H} where $\Delta V$ is the FWHM of the absorption feature.
This reflects that the line is broadened both by the cloud temperature and also by non-thermal processes.

We have produced a publicly accessible data collection containing our processed data \citep{csiro:C3297-2019APRS}. This collection includes the \hi\ cubes, continuum images, source finder catalogues, emission and absorption spectra and catalogues of the gas characteristics.

\subsection{Absorption Towards Source \texorpdfstring{PMN J0029$-$7228}{PMN J0029-7228}}
\label{sec:002919-722811}
\begin{figure}
	\includegraphics[width=\columnwidth]{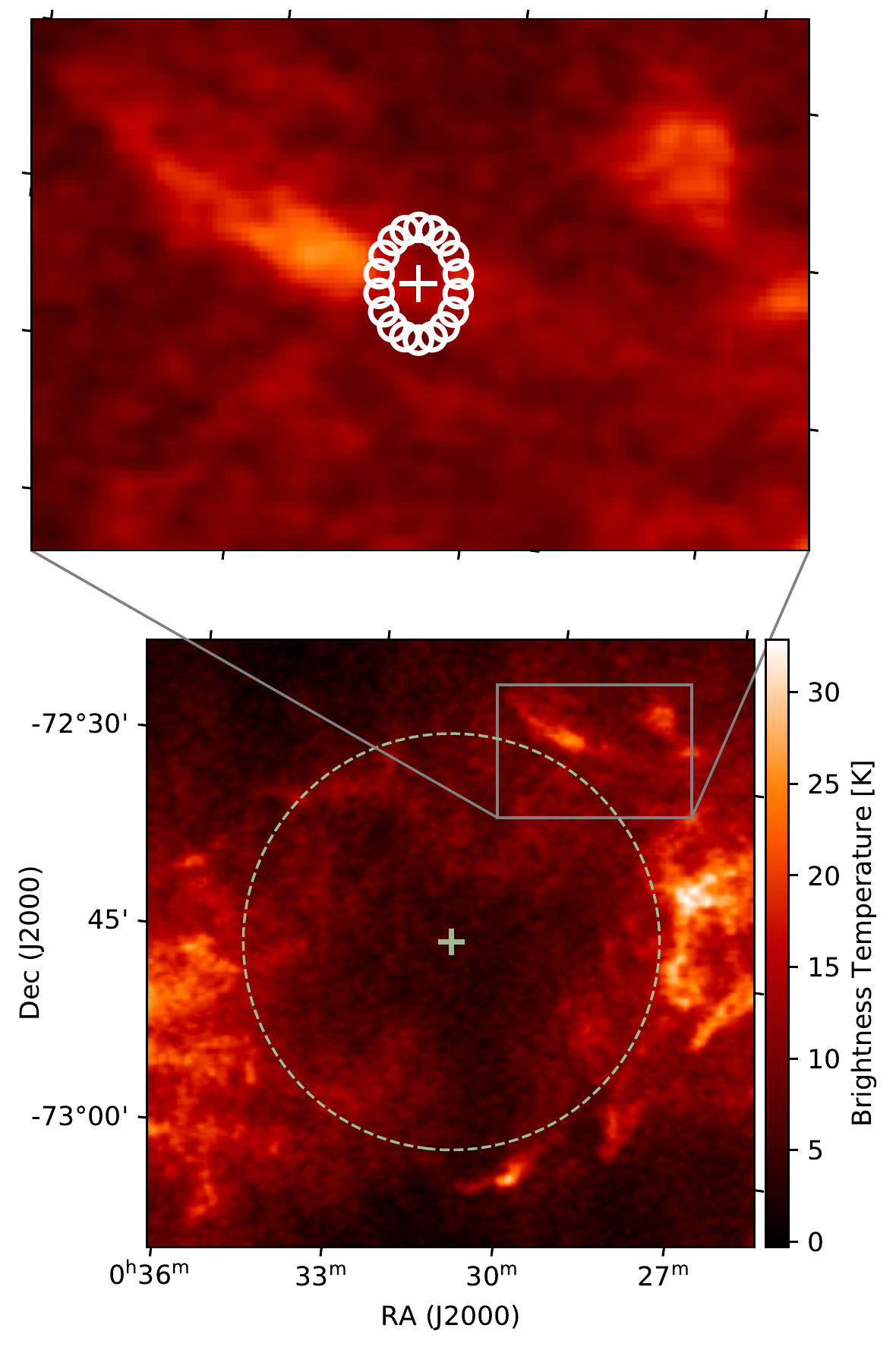}
    \caption{The region around source PMN J0029$-$7228, with the immediate region around the source shown in greater detail. The background is the brightness temperature of the emission at $v_{\rm LSRK,em} = 124.6$ km s$^{-1}$, as measured by the ASKAP SMC observations. In the bottom plot, the shell is shown as a dashed circle, with its centre marked with a plus. In the zoomed plot (top), the location of the emission samples are shown as white circles and source PMN~J0029$-$7228 is shown as a plus. Significant small scale structure is apparent, including a prominent cloud which is likely to be part of the shell wall.}
    \label{fig:emission-day3}
\end{figure}

\begin{figure}
	\includegraphics[width=\columnwidth]{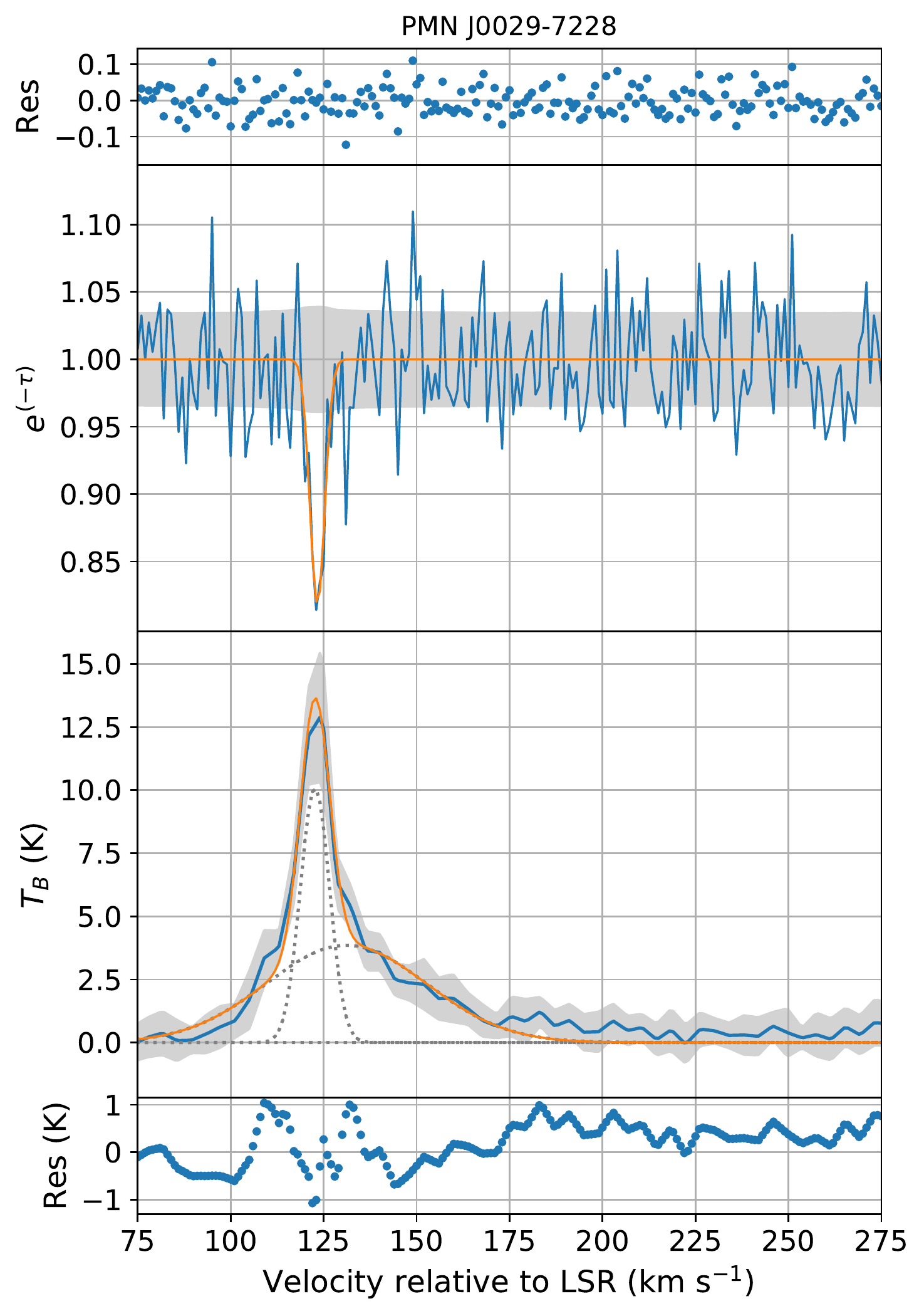}
    \caption{Absorption and emission spectra for the source PMN~J0029$-$7228 showing the velocity range encompassing the SMC emission. The top panel shows the residual between the fitted Gaussian and the observed absorption spectrum. 
    The second panel shows the observed spectrum in blue, the 1$\sigma$ noise level as a grey band and the Gaussian fit in orange. 
    The third panel shows the mean emission taken from ASKAP+Parkes observations in blue, with the 1$\sigma$ noise level as a grey band. The sum of the fitted Gaussians is shown in orange with the individual Gaussians shown as dotted grey lines.
    The bottom panel shows the residual between the fitted Gaussians and the observed emission spectrum.}
    \label{fig:spectrum-day3}
\end{figure}

The \hi\ absorption spectrum for source PMN~J0029$-$7228 is shown in Fig.~\ref{fig:spectrum-day3}.
This is the strongest absorption detection, with 4 consecutive 1~km~s$^{-1}$ channels with at least $3\sigma$ significance.
The single component visible in absorption at $v_{\rm LSRK,abs} = 123.3 \pm 0.3$ km s$^{-1}$ has a FWHM of $4.7 \pm 0.7$ km s$^{-1}$, and $\tau = 0.21 \pm 0.048$.

An image of the emission in this region, as measured by commissioning observations of the SMC with ASKAP \citep{2018NatAs...2..901M} is shown in Fig.~\ref{fig:emission-day3}.
These data have been continuum subtracted. 
Significant small angular scale variation is visible in these higher spatial resolution data.
This is in contrast to the GASS data in which this region is covered by less than two pixels.
The absorption is likely associated with the isolated cloud structure that is prominent in Fig.~\ref{fig:emission-day3}.
The cloud is at a similar velocity to the SMC at this Right Ascension but detached from the main body of the SMC. 
The increased noise level near the emission peak in Fig. \ref{fig:spectrum-day3} is a result of the samples being taken both on and off the cloud and reflects the uncertainty in the emission at the source.
The emission spectrum consists of a cooler narrow component surrounded by a wider and thus warmer component.
The mean column density of the region is $6.3 \times 10^{20}$ cm$^2$. 

Matching the absorption component ($v_{\rm LSRK,abs} = 123.3 \pm 0.3$ km s$^{-1}$) to the nearest emission component ($v_{\rm LSRK,em} = 122.7 \pm 0.2$ km s$^{-1}$, $FWHM_{\rm em} = 9.3 \pm 0.5$ km s$^{-1}$) for source PMN~J0029$-$7228, we find an optical-depth weighted spin temperature (Eq. \ref{eq:spin-temp}) for the absorption component of $T_{\rm s} = 75.2 \pm 22.5$K.  
The absorption component has a $T_{\rm k,max} = 483 \pm 144$ K and an isothermal spin temperature $T_{\rm s,min} = 68 \pm 20$ K (Eq. \ref{eq:spin-temp-est}). 
The kinetic temperature more than six times the spin temperature suggests a highly turbulent environment.
The consistency of the radial velocities of the peaks in emission and absorption indicate that the cool and warm \hi\ are likely to be associated.

\subsection{Absorption Towards Source \texorpdfstring{SUMSS~J012734$-$713639A/B}{SUMSS J012734-713639A/B}}
\label{sec:012734-713638}

\begin{figure}
	\includegraphics[width=\columnwidth]{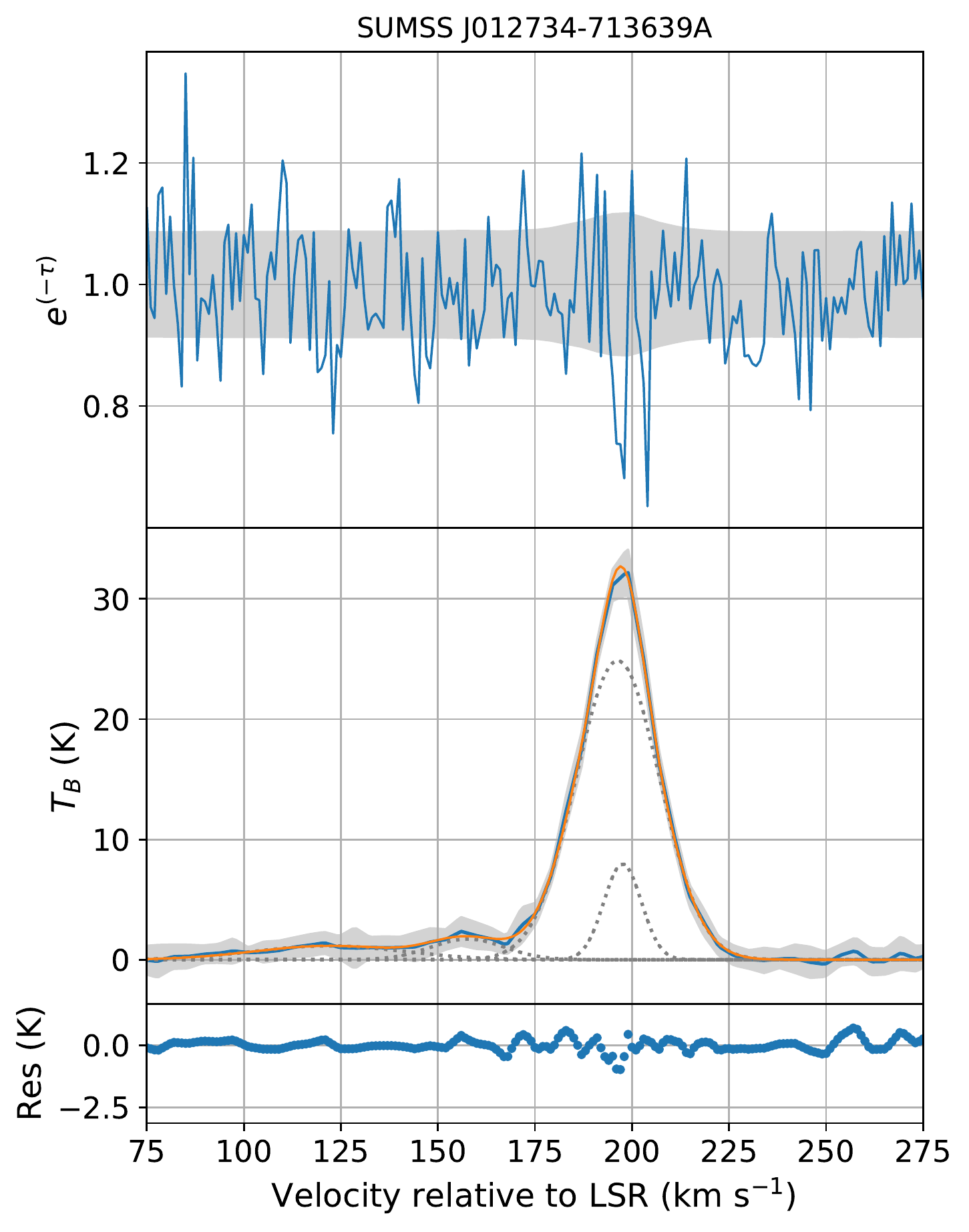}
    \caption{Absorption and emission spectra for the source SUMSS~J012734$-$713639A showing the velocity range encompassing the SMC emission.  
    The top panel shows the observed spectrum in blue, and the 1$\sigma$ noise level as a grey band. 
    The second panel shows the mean emission taken from ASKAP observations in blue, with the 1$\sigma$ noise level as a grey band. The sum of the fitted Gaussians is shown in orange with the individual Gaussians shown as dotted grey lines.
    The bottom panel shows the residual between the fitted Gaussians and the observed emission spectrum.}
    \label{fig:spectrum-day5}
\end{figure}

\begin{figure}
	\includegraphics[width=\columnwidth]{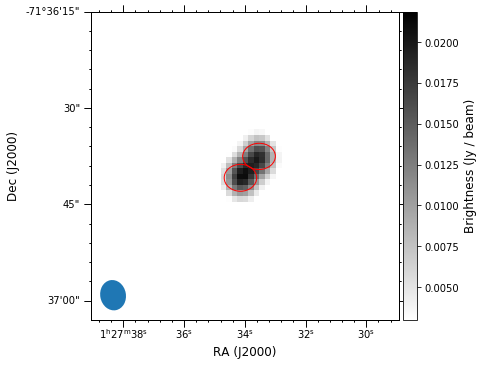}
    \caption{Plot of the inner 45 arcsec of the MFS continuum image surrounding the sources SUMSS~J012734$-$713639A/B. The source ellipses identified by \textsc{Aegean} are shown in red.}
    \label{fig:mfs-day5}
\end{figure}

The spectrum for source SUMSS~J012734$-$713639A is shown in Fig.~\ref{fig:spectrum-day5}.
While there is a clear absorption feature visible at $v_{\rm LSRK,abs} = 198$ km s$^{-1}$, 
there are only 3 consecutive 1~km~s$^{-1}$ channels with at least $2\sigma$ significance
As a result the feature was not fitted with a Gaussian component.
The feature has a width of $\sim5$ km s$^{-1}$, and a maximum $\tau = 0.39 \pm 0.17$.

This field has the highest column density ($N_{\rm \hi} = 15.8\times 10^{20}$ cm$^{-2}$) of the observed fields.
There are variations in column density on small angular scales consistent with noise, but no coherent structures are apparent.
As shown in Fig.~\ref{fig:spectrum-day5}, the velocities of the peaks in the absorption and emission spectra are very close, indicating that the gas being sampled is likely to be associated.

The absorption feature has $T_{\rm k,max} = 546 \pm 218$ K and an isothermal spin temperature $T_{\rm s,min} = 92 \pm 40$ K (Eq. \ref{eq:spin-temp-est}).

The two spatial components of this elongated source (see Fig.~\ref{fig:mfs-day5}) have similar absorption characteristics, so we have presented only the more significant one here.

\section{Discussion}
\label{sec:discussion}

\subsection{Environment of the Cold Gas}

The spectrum in the direction of source PMN~J0029$-$7228 probes a cloud on the very edge of the SMC. 
The cloud, as seen in ASKAP+Parkes emission data, is elongated and extends to either side of the background source.
Note that there appears to be a gap in the emission data at the point of the source (see Fig \ref{fig:emission-day3}) as the ASKAP emission data have had continuum emission removed.
Looking at the surrounding region (see bottom of Fig.~\ref{fig:emission-day3}) we can see that the sampled cloud is part of a shell-like structure.
This structure is centred on $(\alpha, \delta)$ = (00:31:21, -72:44:04) with a radius $\approx$ 16' ($\approx 284$ pc at a distance of 61 kpc; \citealt{2014ApJ...780...59G}).
It has not previously been catalogued as it is outside the range of previous studies \citep[e.g.][]{1997MNRAS.289..225S}.

The SMC is characterised by hundreds of shells with a wide range of scales \citep{1997MNRAS.289..225S}.
Shells are typically circular or elliptical structures of \hi\ swept up by winds from supernovae or massive stars, or from gravitational instabilities \citep{2000ApJ...540..797W}.
We examined potential formation mechanisms for this shell, including examining star formation using H$\alpha$ as a tracer.
No features, however, are apparent in the MCELS H$\alpha$ emission map in the vicinity of the shell \citep{2015ASPC..491..343W}.
There also no known blue or yellow supergiant stars within the shell bounds \citep{2019A&A...629A..91Y}.
However, as noted by \cite{2005MNRAS.360.1171H}, many shells in the SMC, including many in the north-western outer regions, cannot be associated with stellar objects.

We propose that the sampled cloud is a fragment of a large shell and that, with the expansion of the shell, the cloud is being ejected from the SMC at an angle of -300\degr , towards the MS interface region to the north of the SMC \citep{2005A&A...432...45B}.
Based on the SMC distance of $61$ kpc \citep{2014ApJ...780...59G}, the shell fragment has a width of 35 pc.
Such small fragments will eventually evaporate in the ionised medium surrounding the SMC \citep{2019ApJ...887...16S}.
The lifetime can be estimated using $t_{\rm evap} \sim 3.3\times10^{20} n_{\rm c} R_{\rm pc}^2 T_{\rm f}^{-5/2}$ yr \citep{1977ApJ...211..135C}, where $n_{\rm c}$ is the \hi\ density in cm$^{-3}$, $R_{\rm pc}$ is the fragment radius in parsecs and $T_{\rm f}$ is the temperature, in Kelvin, of the surrounding gas.
Adopting a \hi\ density range of 1-10 cm$^{-3}$ \citep{2009ARA&A..47...27K} and a temperature for the hot ionised gas in the surrounding circumgalactic medium of $1\times10^6$ K \citep{2008ApJ...680..276S}, we estimate a fragment lifetime in the range of $10^8$ to $10^9$ years.

Given its location on the edge of the SMC, and its significant longevity, it is possible that this shell fragment was ejected from the SMC as a cloud with a cold core.
Similar outflows have been detected previously, but not direct detections of cold \hi\ gas.
\cite{2018NatAs...2..901M} found a series of outflows of \hi\ from the north of the SMC. 
They found a mixture of shell fragments, filaments and head--tail clouds with radial velocities and spatial separation indicative of the gas leaving the SMC. 
They further demonstrated that some of the gas in the northwest of the SMC was moving away from the SMC faster than the SMC escape velocity of $\approx 85$ km s$^{-1}$ and thus would not fall back into the galaxy.
\cite{2018ApJ...863...49B} demonstrated that even for gas which did not reach escape velocity, ram pressure stripping from the hot MW halo can sweep the gas up and prevent it falling back into the source galaxy.
\cite{2010ApJ...721L..97B,2012MNRAS.421.2109B} showed that outflows from the SMC into the MS can be reproduced using tidal forces from the SMC's interaction with the LMC.
Looking to other gas phases, a potential galactic fountain has been suggested by \cite{2002ApJ...569..233H} to explain the kinematics they observed of hot gas seen in O~\textsc{VI} absorption.
\cite{2015ASPC..491..343W} also noted faint filaments of H$\alpha$ stretching to the north of the SMC towards the MS interface region.
From the northwest of the SMC this cloud would flow into the MS interface region and then be swept into the MS \citep{2005A&A...432...45B}.
When compared to the MS age of $\sim2.5\times 10^9$ yr \citep{2010ApJ...723.1618N}, this indicates that similar clouds could be found well into the MS.

\subsection{Environment of the Non-detections}

The three fields in which we did not detect cold \hi\ are shown as dotted red circles in Fig.  \ref{fig:sources}.
These fields contain the sources
004327$-$704136A/B, 004938$-$700133 and 010458$-$703735.
They are all further away from the SMC than the two detections and they also have lower column densities.
As noted in Table \ref{tab:sources_table}, these sources have column densities in the SMC velocity range ($75 \leq v_{\rm LSRK} \leq 275$) ranging from half to a fifth of the PMN~J0029$-$7228 detection.
Additionally, in the high spatial resolution ASKAP data, they do not show any coherent structures apart from noise-like variations.

Using emission data, \cite{2018NatAs...2..901M} found cold \hi\ outflow features in the SMC in including a 'hook' feature (see their Fig.~2).
The source PMN~J0104$-$7038 is $\approx 15$ arcmin in RA from this hook feature, many times the width of the feature.
So while the source is close to the feature it does not sample the gas of the feature.
A fainter background source, SUMSS~J010622$-$704146, closer to the feature, was included in the same field but was well away from the beam centre. 
As a result it was too faint (9 mJy) in our observations to get significant absorption detections.
However, future ASKAP absorption observations, with its more uniform sensitivity, will be able to gain better data on this source.

\subsection{Comparison to the Magellanic Stream}

\cite{2009ApJ...691L.115M} detected cold \hi\ in the MS in front of the source J0119$-$6809, as shown in Fig. \ref{fig:sources}.
They identified two \hi\ absorption spectral components with $\tau$ = 0.044 and $\tau$ = 0.052 against the 659 mJy source.
The spin temperatures of the components were measured as $79 \pm 9$ K and $68 \pm 6$ K respectively.
These spin temperatures are within $1\sigma$ of the $68 \pm 20$ K we measured towards source PMN~J0029$-$7228, our best detection.
They have lower uncertainties due to the much brighter background source (659 mJy versus our 88 mJy source).
With a column density of $2.3\times 10^{20}$ cm$^{-2}$, this gas is in a region with almost half the column density as for source PMN~J0029$-$7228.
The \cite{2009ApJ...691L.115M} detection is in a region with small scale structure, in this case a small cloud with higher density than surrounding gas, similar to our detection. 

Indirect detections of cold \hi\ in emission have also been made in the MS.
The first measurement of cool \hi\ in the MS was made by \cite{2005A&A...432...45B}, who identified an \hi\ spectral line component in a clump of dense gas with a FWHM line width of $\approx $4 km s$^{-1}$, indicating $T_{\rm k} \approx 350$ K.
In an analysis of \hi\ emission data from the Leiden/Argentine/Bonn survey (LAB; \citealt{2005A&A...440..775K}) for the MS, \cite{2006A&A...455..481K} found that 27\% of the \hi\ mass of the MS within 50\degr\ of the LMC was in 
cold cores, regions of narrow line width \hi\ components surrounded by wider and thus warmer components.
In the portion of the MS more than 50\degr\ from the centre of the LMC, only 10\% of \hi\ was in cold cores.
\cite{2008ApJ...680..276S} surveyed the MS tip (in the range 90--120\degr\ from the LMC) at 3.5' resolution with the Arecibo telescope, which enabled them to resolve individual \hi\ clouds which would have been unresolved and beam-smeared in lower resolution earlier surveys (e.g. the LAB survey).
They observed that 12\% of sight-lines in the MS tip had \hi\ clouds with cold cores and warmer envelopes.
The cold cores had FWHM line widths ranging from 3---20~km~s$^{-1}$.
The emission spectra of our two detections also exhibited this structure of a narrow, cooler component surrounded by a wider, warmer component, although the cooler component was more pronounced in our better detection.
These results build up a picture of cold \hi\ gas in the MS occurring in dense cores shielded by surrounding warmer \hi\ envelopes.

These clouds with cold cores are analogous to the high velocity clouds (HVCs) observed in the Leading Arm. 
\cite{2006A&A...457..917B} discuss two head--tail clouds, one with both narrow and wide components in the head and the other with only narrow components. 
They find that the cool core, represented by the narrow components, can survive while the outer layer of warmer \hi\ is being stripped away by ram pressure stripping by the hot MW halo.
\cite{2016MNRAS.461..892F} examined five HVCs and found that the resolved clouds all had cold cores and thermal pressures consistent with a two-phase equilibrium in SMC metallicities.
Looking back to the MS, \cite{2003ApJ...586..170P} found that similar head--tail clouds were prevalent across the Stream, with many elongated along the MS. 
Unfortunately, the velocity resolution (15 km/s) of the \cite{2003ApJ...586..170P} emission data made it impossible to know whether the head--tail clouds had cold gas.
\cite{2014ApJ...792...43F} found that 28\% of high velocity clouds in the MS had a head--tail structure.

Whilst emission observations have provided this picture of cold \hi\ gas in the MS, we only have three direct detections, including the two from this study.
To measure the spin temperatures of the cold \hi\ gas across the extent of the MS and to provide a comparison for the emission data, we need far more absorption measurements. 
The wide field of view of ASKAP, combined with high spatial and spectral resolution of the Galactic ASKAP (GASKAP) survey \citep{2013PASA...30....3D} provide us with an opportunity to address this need.

\subsection{Future prospects for cold gas detection}

We have shown in this study that regions with more small scale emission, despite lower column density, had clearer cold gas detections.
As shown with the absorption detected against source SUMSS~J012734$-$713639A/B, even faint background sources (18--20 mJy) can highlight the presence of cold gas.
In previous targeted surveys of the Magellanic system \citep[e.g.][]{2019ApJS..244....7J,2009ApJ...691L.115M}, sources for observations were selected based on regions of higher \hi\ density and bright extra-galactic continuum sources.
For example, \cite{2019ApJS..244....7J} selected sources with flux density $S_{\rm cont}$ > 50 mJy behind areas of \hi\ column density of $N_{\rm \hi}$ $\geq 5 \times 10^{20}$ cm$^{-2}$.
As this study has shown, observations of the lower density gas ($6-16\times10^{20}$ cm$^{-2}$) against dimmer background sources ($18-88$ mJy) are viable.
This opens up many more background sources as targets against which we might feasibly detect cold \hi\ gas in the MS.

The time required to obtain low noise absorption spectra has been the main limiting factor in building up a dataset of direct cold gas observations.
By combining a wide field of view with high spectral and spatial resolution, ASKAP is uniquely placed to allow wide-spread \hi\ absorption observations of the entire Magellanic system.
The GASKAP observing strategy of long duration observations of each field will maximize this potential.
The predicted noise levels for the GASKAP observations of the MS are $\sigma_F$ = 2.0 mJy for 5 kHz (1 km s$^{-1}$) channels \citep{2013PASA...30....3D}, a 60\% improvement on the noise levels in this study. 
Thus GASKAP will be well placed to examine this regime of dimmer extra-galactic sources shining through lower density gas in the MS.

\section{Conclusions}
\label{sec:conclusions}

We have presented the results of high spectral and spatial resolution, deep observations with the ATCA looking for cold \hi\ in absorption against five extra-galactic continuum sources in the MS interface region on the outskirts of the SMC.
These sources were selected from sources identified by  \cite{Buckland-Willis:anu} that showed candidate areas of higher opacity.

We have detected cold \hi\ in two of the five locations.
These detections were against the strongest (88 mJy) and weakest (18 mJy) of the primary target sources.
The detections were in the two highest column density regions ($6-16\times10^{20}$ cm$^{-2}$) observed, although all regions are well below typical SMC column densities.

The clearest detection, at source PMN~J0029$-$7228, is associated with small scale structure, in this case a cloud forming a fragment of a shell on the outskirts of the SMC.
The fragment is positioned on the outside of the SMC from the centre of the shell, making it a candidate for breaking out of the SMC as the shell expands. 
Once outside of the SMC the fragment would be swept up by ram pressure stripping by the hot Milky Way halo and into the MS.
Despite being embedded in the ionised halo surrounding the SMC, the size of the fragment is such that it will last for hundreds of millions of years and potentially reach the MS.
Based on this example, we suggest that breaking super shells from the Magellanic Clouds may be a source of cold \hi\ gas to supply the rest of the Magellanic system.

Finally we find that the regime of low column density \hi\ lit by faint extra-galactic sources is a viable source for cold gas detections in the MS.
The GASKAP survey of the MS, with its high spatial and spectral resolution, coupled with long dwell times to improve sensitivity and untargeted surveying of absorption, will be very well suited to explore this regime.

\section*{Acknowledgements}

The Australia Telescope Compact Array and Parkes radio telescope are part of the Australia Telescope National Facility which is funded by the Australian Government for operation as a National Facility managed by CSIRO.
The Australian SKA Pathfinder is part of the Australia Telescope National Facility which is managed by CSIRO. Operation of ASKAP is funded by the Australian Government with support from the National Collaborative Research Infrastructure Strategy. ASKAP uses the resources of the Pawsey Supercomputing Centre. Establishment of ASKAP, the Murchison Radio-astronomy Observatory and the Pawsey Supercomputing Centre are initiatives of the Australian Government, with support from the Government of Western Australia and the Science and Industry Endowment Fund. We acknowledge the Wajarri Yamatji people as the traditional owners of the Observatory site.
N.Mc.-G. and K.J. acknowledge funding from the Australian Research Council via grant FT150100024. 
This research has made use of the SIMBAD database, operated at CDS, Strasbourg, France.
This research has made use of the NASA/IPAC Extragalactic Database (NED), which is funded by the National Aeronautics and Space Administration and operated by the California Institute of Technology.
We thank the reviewer, Kat Barger, for thoughtful comments and suggestions which have improved this work.




\bibliographystyle{mnras}
\bibliography{cold_gas_smc}







\bsp	
\label{lastpage}
\end{document}